# Image of optical diffraction radiation (ODR) source and spatial resolution of ODR beam profile monitor.

A.P. Potylitsyn

*Tomsk Polytechnic University, Russian Federation*

**Abstract**

The approach to obtain the image shape of an optical diffraction radiation (ODR) source focused by lens on a detector with taking into account the "pre-wave zone" effect has been developed. In full analogy with optical transition radiation source the characteristic size of ODR slit image doesn't depend on the Lorentz-factor and is defined by slit width, wavelength and lens aperture.

If a beam is passed through a narrow slit the shape of a slit image depends on a beam size. ODR generated by extremely ultrarelativistic beam passing through a narrow slit may be used to determine the beam size from ODR slit image.

**1.** In the work [1] the authors have developed optical transition radiation (OTR) beam profile monitor to measure a transverse size of an electron beam with the resolution $\sigma \sim 2$ mcm. The theoretical works [2-4] based on the wave optics have shown that the size of OTR source image (OTR generated by single dimensionless particle) is determined by the amplification of the optical system $M$, the angular aperture of focusing lens $\theta_0$ and OTR wavelength $\lambda$:

$$\sigma \sim \frac{\lambda}{\theta_0} M . \tag{1}$$

For real optical systems ($M = 1$, $\theta_{0\,max} \sim 0.2$) the minimal resolution $\sigma_{min}$ runs into the value of a few wavelengths. The similar spatial resolution is sufficient for the diagnostics of the beams with the transverse sizes of the order of 5÷10 mcm.

But as it has been mentioned in [1] the interaction of the beam with small transverse sizes with target may lead to a changing of optical characteristics of OTR target after the passing of a few bunches with the population $\sim 10^{10}$ $e^-$/bunch. That is why the methods of noninvasive diagnostics, where a beam doesn't interact with a target directly are developing rather intensively in recent years [5]. The methods based on the use of optical diffraction radiation (ODR) are the latest ones [6]. The spatial resolution of ODR beam profile monitor at the KEK-ATF extracted electron beam with the Lorentz-factor $\gamma = 2500$ has been investigated in the cited work. The higher the energy of the beam particle is, the bigger the "natural" size of the luminescent area on the target, which is determined by the parameter $\gamma\lambda$. Thus, the distance from the target $a$, where the similar OTR source with the divergence $\sim \gamma^{-1}$ can be observed as pointwise, increases proportionally to the square of the Lorentz-factor [7]:

$$a \gg L_0 = \gamma^2 \lambda . \tag{2}$$

In other words, all characteristics of optical devices (including image size) set at the distances less than $L_0$, (that is, they are situated in pre-wave zone following the terminology [7]) should be calculated with regard to the finite sizes of the luminescent spot.

For example, in case of SLAC FFTB beam with $\gamma = 60000$, the estimate (2) gives $L_0 \sim 1.8$ km for wavelength $\lambda = 0.5$ mcm. So, it is clear, that for a reasonable distance between a target and an optical system ($a \sim 2$ m) the characteristic relationship will be:

$$R = \frac{a}{L_0} = \frac{a}{\gamma^2 \lambda} \sim 10^{-3} \ll 1,$$

that is, the monitor is situated in the extremely pre-wave zone.

**2.** Let us consider the image of OTR source with regard to the "pre-wave zone" effect. For simplification of calculations let us consider backward transition radiation (BTR) for perpendicular particle passing through the target. In real conditions the inclination of a target $\psi$ relative to the particle trajectory differs from 90°, but, as it has been shown in [8,9], the angular characteristics of BTR of ultrarelativistic particles in a wave zone (backward diffraction radiation, BDR, as well) are determined relative to the mirror reflection direction and don't depend on the slope of the target $\psi$, if the relationship $\psi \gg \gamma^{-1}$ is fulfilled. Following [3], let us write the expression for TR field components generated by ultrarelativistic particle in an infinite perfect target $L$ with the focus distance $f$ on the detector $D$ (the distance between the target and the lens stands $b$, between the lens and the detector stands $a$, the standard relationship $\frac{1}{a} + \frac{1}{b} = \frac{1}{f}$ is fulfilled, see Fig. 1).

$$\begin{aligned}
E_{x,y}^D (\vec{r}_T) &= const \int d\vec{R}_T \int d\vec{R}_L \begin{Bmatrix} \cos\varphi_S \\ \sin\varphi_S \end{Bmatrix} \frac{K_1(kR_T/\beta\gamma)}{\beta\gamma} \times \\
&\times \exp\left[i\frac{k}{2a}R_T^2\right] \exp\left[i\frac{k}{2a}R_L^2\right] \exp\left[-i\frac{k}{a}R_L R_T \cos(\varphi_S - \varphi_L)\right] \times \\
&\times \exp\left[-\frac{ik}{2f}R_L^2\right] \exp\left[i\frac{k}{2b}(\vec{R}_L - \vec{R}_D)^2\right].
\end{aligned} \quad (3)$$

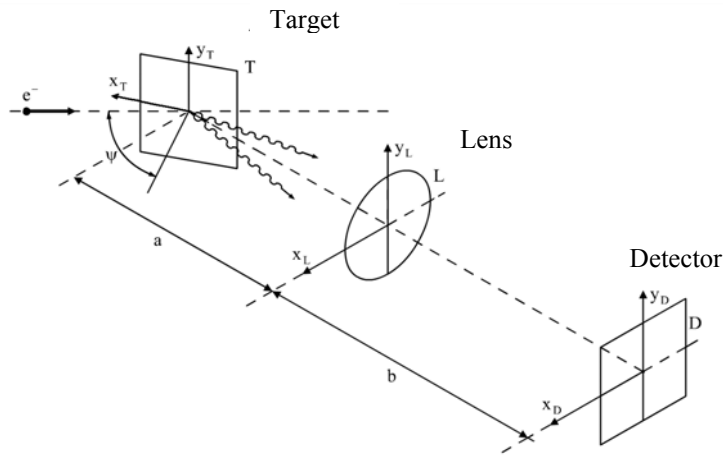

In (3) the coordinates are indicated by the indexes $T$, $L$, $D$ in the system on target, lens and detector surfaces subsequently (see Fig.1), $K_1$ – modified Bessel function, $k = \frac{2\pi}{\lambda}$ – wave number.

Using the known Bessel's function representation and introducing dimensionless variables

Fig. 1. Optical system scheme of OTR beam size monitor.

$$r_T = \frac{2\pi R_T}{\gamma\lambda}, \quad r_L = \frac{\gamma}{a} R_L, \quad r_D = \frac{2\pi R_D}{\gamma\lambda} \qquad (4)$$

it is possible to get the expression for the radial component of the transition radiation field for the infinite boundary target:

$$E_D(r_D, R) = const \int r_T dr_T \int r_L dr_L \, K_1(r_T) J_1(r_T r_L) \exp\left[i\frac{r_T^2}{4\pi R}\right] \times J_1\left(r_L \frac{r_D}{M}\right). \qquad (5)$$

Hereafter, we shall make calculations for $M = 1$ (that is, $b = a = 2f$).
In a wave zone, that is, at the condition fulfilment
$$R \gg 1$$
the exponent can be replaced by unit. In this case the integration over $r_T$ is carried out analytically:

$$E_\infty(r_L) = \int_0^\infty r_T \, dr_T \, K_1(r_T) J_1(r_T r_L) = \frac{r_L}{1 + r_L^2}, \qquad (6)$$

and for the field (5) we shall get the following formula

$$E_D(r_D) = const \int_0^{r_m} r_L \, dr_L \, \frac{r_L}{1 + r_L^2} J_1(r_L r_D),$$

which coincides with the result obtained earlier [3]. The calculation of OTR field in pre-wave zone (for example, on the lens surface) can be performed only numerically:

$$E_L(r_L, R) = const \int_0^\infty r_T \, dr_T K_1(r_T) J_1(r_T r_L) \exp\left[i\frac{r_T^2}{4\pi R}\right]. \qquad (7)$$

For the calculation simplification we shall use approximation
$$r_T K_1(r_T) = (1 + 0.57 r_T - 0.04 r_T^2) e^{-r_T}, \qquad (8)$$

which will give an error in some percent in the interval $0 \le r_T \le 5$ (see Fig. 2).
When calculating the integral $r_T$ the integration is carried out in the limits for simplicity $0 \le r_T \le 5$.

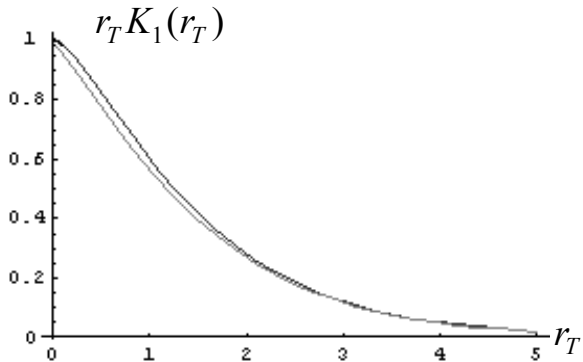

Fig. 2. Approximation of the dependence $r_T K_1(r_T)$ - upper curve, by the formula (8) – lower curve.

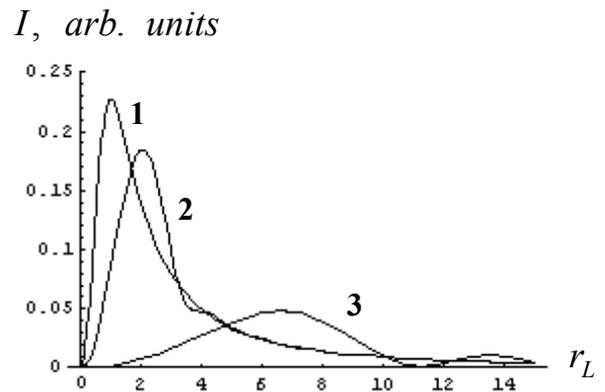

Fig. 3. Radial distribution of OTR intensity on the lens surface for different parameter R values (Curve 1 – $R = 10/2\pi$; 2 – $R = 1/2\pi$; 3 – $R = 0.1/2\pi$).

The intensity dependence $I = |E_L(r_L, R)|^2$ are shown in Fig. 3 for different values $R$, which are in good agreement with V. Verzilov's calculations (see Fig. 3 in article [7]).

The OTR intensity on the lens $|E_L(r_L R)|^2$ achieves maximum value at such radius $r_L = r_0$, when the maximal "overlapping" of oscillating subintegral functions $J_1(r_T r_L)$ and $Re\left[\exp\left(i\frac{r_T^2}{4\pi R}\right)\right]$ occurs. As far as the main contribution to the integral (5) will give the range $r_T \leq 1$, this requirement comes to the requirement of the coincidence of the first nulls of the both functions. Thus, we have two relationships

$$r_T r_0 = 3.832,$$
$$\frac{r_T}{4\pi R} = \frac{\pi}{2}, \quad (9)$$

from which we can find the connection between $r_0$ and $R$:

$$r_0 = \frac{3.832}{\sqrt{2}\,\pi\sqrt{R}} = \frac{0.863}{\sqrt{R}}. \quad (10)$$

Returning to the size variables, we can get:

$$R_L = 0.863\sqrt{a\lambda}. \quad (11)$$

The estimate (10) for $R = \frac{1}{2\pi};\ \frac{0.1}{2\pi}$ gives the values $r_0 = 2.16$ and $6.84$, which agree reasonable with the exact values $r_0 = 2.05$ and $6.65$ (see Fig.3). It follows from the first equation (9) that in the pre-wave zone the target range gives the main contribution in OTR intensity

$$R_T \leq \sqrt{\frac{a\lambda}{2}}, \quad (12)$$

that is, the estimates (11), (12) don't depend on the Lorentz-factor in the case, which is considered.

To obtain the image field (5) on the detector, it is necessary to calculate a double integral. It should be noticed that the "inner" integral in (5) is taken analytically:

$$G(r_T, r_D, r_m) = \int_0^{r_m} r_L dr_L\, J_1(r_T r_L) J_1(r_L r_D) =$$
$$= \frac{r_m}{r_D^2 - r_T^2}\{r_T J_0(r_m r_T) J_1(r_m r_D) - r_D J_0(r_m r_D) J_1(r_m r_T)\}. \quad (13)$$

Here, angular lens aperture is defined as $r_m$ (in the angle units $\gamma^{-1}$). Thus, the field on the detector is calculated through the single integral:

$$E_D(r_D, R) = \int r_T dr_T\, K_1(r_T) G(r_T, r_D, r_m) \exp\left[i\frac{r_T^2}{4\pi R}\right] =$$
$$\int_0^5 dr_T (1 + 0{,}57 r_T - 0{,}04 r_T^2) \exp\left[-r_T + i\frac{r_T^2}{4\pi R}\right] G(r_T, r_D, r_m) \quad (14)$$

In Fig. 4 normalized shapes of OTR images on the detector $I = |E_D(r_L, R)|^2$ are shown for $r_m = 50, 100$ in wave zone ($R = 1$).

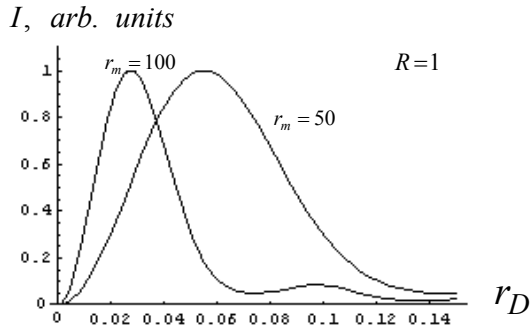

Fig. 4. Normalized shape of OTR source image on the detector plane for lens aperture $r_m = 100$ (left curve) and $r_m = 50$ (right curve) in wave zone ($R = 1$).

For the Lorentz-factor $\gamma = 1000$ the indicated lens aperture values correspond to the angles $\theta_m = 0.05$ and $\theta_m = 0.1$, for which the quantity $|E_D|^2$ has been calculated in [3] (see Fig. 4).

The results obtained by different methods coincide with a good accuracy. Thus, for example, for aperture $\theta_m = 0.1$ rad the maximum in the radial distribution corresponds to the radius $\rho_L^{max} = 4.4\lambda$ [3]. As it follows from Fig.4 (left curve) the nondimensional variable value $r_D = 0.0275$ corresponds to the maximum for $r_m = 100$, that for $\gamma = 1000$ corresponds to the radius $R^{max} = \frac{1000}{2\pi} \cdot 0.0275\lambda = 4.38\,\lambda$. At aperture decreasing twice, the radius $r_D^{max}$ increases in two times.

The images in pre-wave zone ($R = 0.001$) calculated by formula (14) are shown in Fig. 5. It can be noticed that the normalized distribution shape on the detector depends on the parameter $R$ very slightly if the lens aperture $\theta_m$ remains constant.

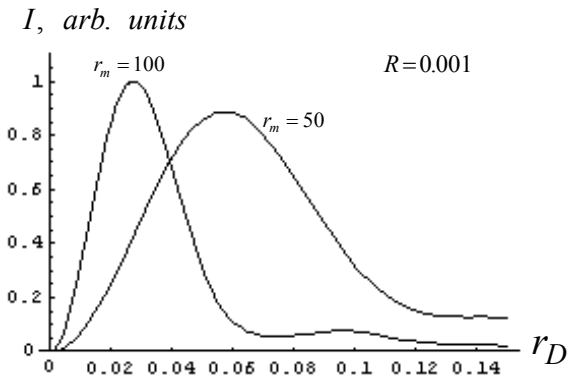

Fig. 5. OTR source image in «pre-wave zone» ($R = 0.001$) for the same conditions and with the same normalizing factors.

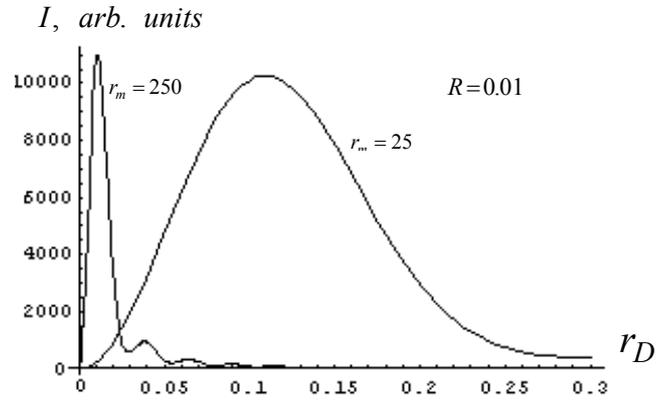

Fig. 6. OTR source image at the fixed lens diameter for different distances $a$ between the target and the lens ($R = 0.01; r_m = 250$ – right curve; $R = 0.1; r_m = 25$ – left).

In a real case, at the change of the distance $a$ between the lens and the target for the same parameters of the task, not only the parameter $R$ changes, but lens aperture $r_m$ at its fixed diameter also.

OTR spot images for $R = 0.1$, $r_m = 25$ and $R = 0.01$, $r_m = 250$ are shown in Fig.6 that corresponds to the fixed lens diameter at the distance change in 10 times. The right curve is multiplied by 100 times for the sake of convenience. As it follows from the figure, the spot size (monitor resolution) is decreased in 10 times at the lens approach to the target (the angular aperture is increased in the same number of times).

**3.** The approach developed in the previous paragraph gives the opportunity to obtain the "image" of a round hole with the radius $\rho_0$, in the center of which the charged particle flies and generates the optical diffraction radiation (ODR) (see [10]). ODR field on the detector is calculated by the formula, which is analogous to (14):

$$E_D^{DR}(r_D, R, \rho_0) = \int_{\rho_0}^{5} dr_T \left(1 + 0.57 r_T - 0.04 r_T^2\right) \exp\left[-r_T + i\frac{r_T^2}{4\pi R}\right] G(r_T, r_D, r_m). \quad (15)$$

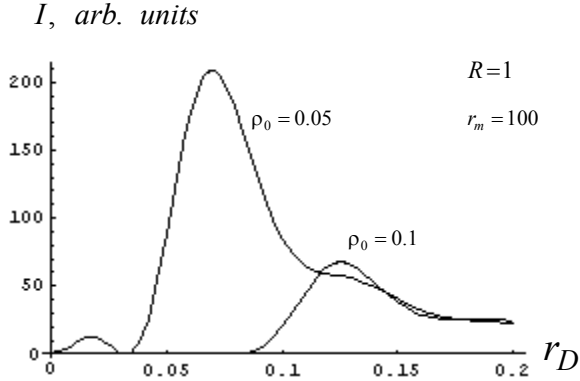

Fig. 7. The image of the round hole in wave zone ($\rho_0 = 0.05$ – left curve; $\rho_0 = 0.1$ right) for aperture $r_m = 100$, $R = 1$.

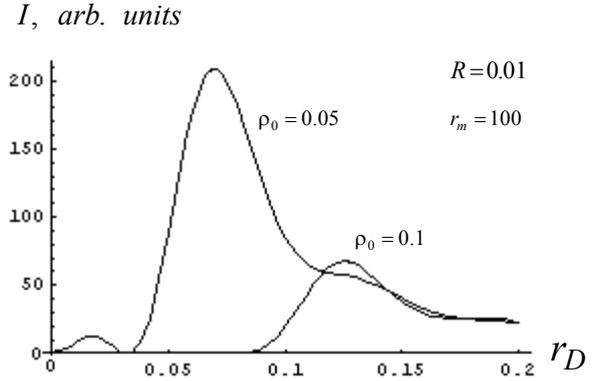

Fig. 8. The image of the round hole in "pre-wave zone" ($\rho_0 = 0.05$ – left curve; $\rho_0 = 0.1$ right) for aperture $r_m = 100$ and $R = 0.01$.

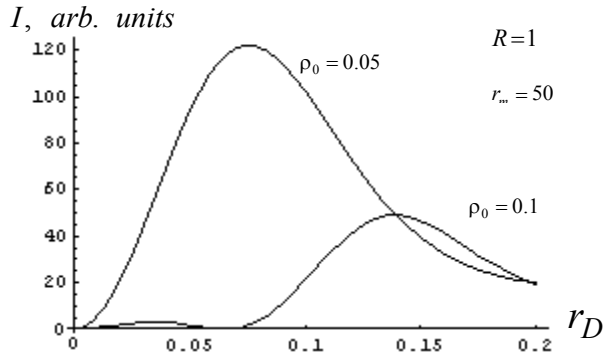

Fig. 9. The hole image ($\rho_0 = 0.05$, $\rho_0 = 0,1$) in wave zone ($R = 1$) with a poor resolution ($r_m = 50$).

Using field (15) the similar "images" of the ODR source (round hole) for $\rho_0 = 0.05$ and $\rho_0 = 0.1$ in wave zone ($R = 1$) were performed (see Fig. 7). If the spatial resolution of OTR (in other words, the image size of ODR from the hole with the infinite small radius) is determined by the quantity $\lambda/\theta_0$ (or in our variables, quantity $\Delta r_D \sim 0.0275$ for $r_m = 100$), it should be expected, that at $\rho_0 < \Delta r_D$ the ODR source image will coincide practically with OTR one, but for $\rho_0 > \Delta r_D$ one may obtain the hole image on the detector. As it follows from the figure 7, the intensity of radiation on the detector at $r_D < \rho_0$ is practically lacking and achieves the maximum at $r_D^m \approx r_T + \Delta r_D^{DR}$, with the maximum width, which is determined by the quantity $\Delta r_D^{DR} \approx 0.04$ in both cases.

The analogous dependences for the pre-wave zone are shown in figure 8. It can be noted that for ODR the pre-wave zone effect reveals slightly in the image plane. But the lens aperture, as in OTR case, influences the width of maxima near hole edge image (see Fig. 9). At aperture decreasing two times (up to $r_m = 50$) the width of the distributions increases more than two times in comparison with $r_m = 100$ (see Fig. 7, 8). In this case the hole image with $\rho_0 = 0.05$ is too smoothed.

Using the formulas (13), (15) we can get ODR characteristics in a pre-wave zone for the disk of finite sizes having substituted the upper limit in integral (15) for the outer disk radius $\rho_{max}$.

In wave zone the use of the formula (5) is the more simple method. We can write instead of the formula (6)

$$E^{DR}(r_L) = \int_{\rho_0}^{\rho_{max}} r_T dr_T\, K_1(r_T) J_1(r_T r_L) = -\frac{r_T}{1+r_L^2}[K_0(r_T)J_1(r_T r_L) + r_L K_1(r_T)J_0(r_T r_L)]\,|_{\rho_0}^{\rho_{max}}. \quad (16)$$

For the hole radius $\rho_0 \ll 1$ with a good accuracy $\rho_0 K_0(\rho_0) \approx 0$, $\rho_0 K_1(\rho_0) \approx 1$, and, consequently, at $\rho_{max} \to \infty$

$$E_\infty^{DR}(r_L) \approx \frac{r_L}{1+r_L^2} J_0(\rho_0 r_L). \quad (17)$$

The formula (17) is in a good agreement with the result obtained earlier [11].
But at the violation of axial symmetry of the task (for example, at the particle flight not through the center of the hole), the formulas obtained above don't work.

**4.** ODR beam size monitor [6] is based on the measuring of the degree of the deformation of ODR angular distribution from the slit in a wave zone generated by electron beam with finite transverse size, in comparison with the distribution from the infinite narrow beam.

To estimate whether it is possible to obtain the information about transverse sizes of the beam passing through the slit using the ODR slit image at the focusing on the detector, is of interest. In this case it is necessary to use Cartesian coordinates in the initial formulas.

As before, we shall introduce Cartesian coordinates on the target, lens and detector using $T$, $L$, $D$ indexes. Dimensionless variables will be introduced analogously (4):

$$\begin{Bmatrix} x_T \\ y_T \end{Bmatrix} = \frac{2\pi}{\gamma\lambda}\begin{Bmatrix} X_T \\ Y_T \end{Bmatrix}, \quad \begin{Bmatrix} x_L \\ y_L \end{Bmatrix} = \frac{\gamma}{a}\begin{Bmatrix} X_L \\ Y_L \end{Bmatrix}, \quad \begin{Bmatrix} x_D \\ y_D \end{Bmatrix} = \frac{2\pi}{\gamma\lambda}\begin{Bmatrix} X_D \\ Y_D \end{Bmatrix}. \quad (18)$$

Thus, instead of (3) we shall have:

$$\begin{Bmatrix} E_x^D(x_D, y_D) \\ E_y^D(x_D, y_D) \end{Bmatrix} = const \int dx_T\, dy_T \int dx_L\, dy_L \begin{Bmatrix} x_T \\ y_T \end{Bmatrix} \frac{K_1\left(\sqrt{x_T^2 + y_T^2}\right)}{\sqrt{x_T^2 + y_T^2}} \times$$

$$\times \exp\left[i\frac{x_T^2 + y_T^2}{4\pi R}\right]\exp[-i(x_T x_L + y_T x_L)]\exp\left[-i\left(x_L \frac{x_D}{M} + y_L \frac{y_D}{M}\right)\right]. \quad (19)$$

In (19), as before, $R = \frac{a}{\gamma^2 \lambda}$; $M$ – magnification. For a rectangular lens:

$$-x_m \leq x_L \leq x_m;\quad -y_m \leq y_L \leq y_m. \quad (20)$$

this integral is calculated very simply. In this case the "inner" double integral is taken analytically again:

$$\int_{-x_m}^{x_m} dx_L \int_{-y_m}^{y_m} dy_L \exp\left[-ix_L\left(x_T + \frac{x_D}{M}\right)\right]\exp\left[-iy_D\left(y_T + \frac{y_D}{M}\right)\right] =$$

$$= 4\frac{\sin\left[x_m\left(x_T + \frac{x_D}{M}\right)\right]}{x_T + \frac{x_D}{M}} \cdot \frac{\sin\left[y_m\left(y_T + \frac{y_D}{M}\right)\right]}{y_T + \frac{y_D}{M}} = G_x(x_T, x_D, x_m) \cdot G_y(y_T, y_D, y_m) \quad (21)$$

Thereafter we shall consider the case $M = 1$ again. Thus the expression (19) is reduced to the double integral over the target surface:

$$\begin{Bmatrix} E_x^D(x_D, y_D) \\ E_y^D(x_D, y_D) \end{Bmatrix} = const \int dx_T dy_T \begin{Bmatrix} x_T \\ y_T \end{Bmatrix} \frac{K_1\left(\sqrt{x_T^2 + y_T^2}\right)}{\sqrt{x_T^2 + y_T^2}} \times$$
$$\times \exp\left[i\frac{x_T^2 + y_T^2}{4\pi R}\right] G_x(x_T, x_D, x_m) \cdot G_y(y_T, y_D, y_m). \quad (22)$$

After integration in (22) in the limits
$$-5 \leq x_T, y_T \leq 5 \quad (23)$$
at $R \gg 1$ the standard images of OTR field components are obtained (see Fig. 10).

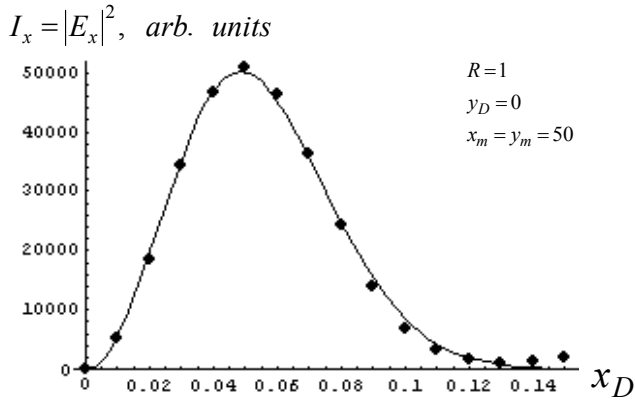

Fig. 10. $X_D$ - component of OTR source intensity in wave zone at $y_D = 0; R = 1; x_m = y_m = 50$ (points).

The fit of the first function maximum is shown in the same figure by line

$$F(x_D) = 5.72 \cdot 10^7 \, x_L^2 \exp\left[-420 \, x_L^2\right].$$

Or in usual variables (for $\gamma = 1000, \quad \Delta\theta_x = 0.05$)

$$F\left(\frac{x_D}{\lambda}\right) = 2260 \left(\frac{x_D}{\lambda}\right)^2 \exp\left[-0.0166\left(\frac{x_D}{\lambda}\right)^2\right].$$

**5.** At particle flight in the center of the slit of width $2h$, directed along the axis $x$, ODR field is calculated by formula (22) at the integration over $y_T$ in the limits $\{-5,-h\}, \{h,5\}$. In Fig. 11 the slit image ($y_d$ - distribution) in wave zone ($R = 1$) for the aperture $x_m = y_m = 50$ is shown.

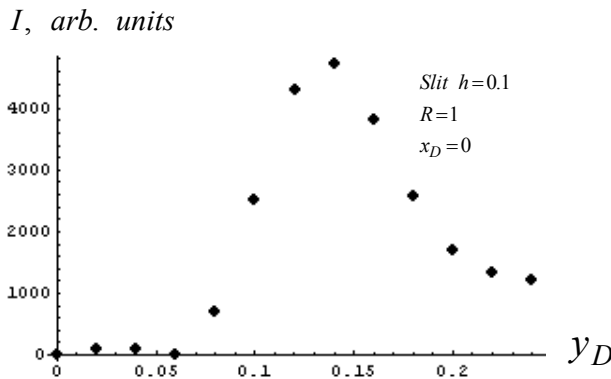

Fig. 11. The ODR slit image for slit width $h = 0.1$ in wave zone for aperture $x_m = y_m = 50$ ($y_D = 0; R = 1$).

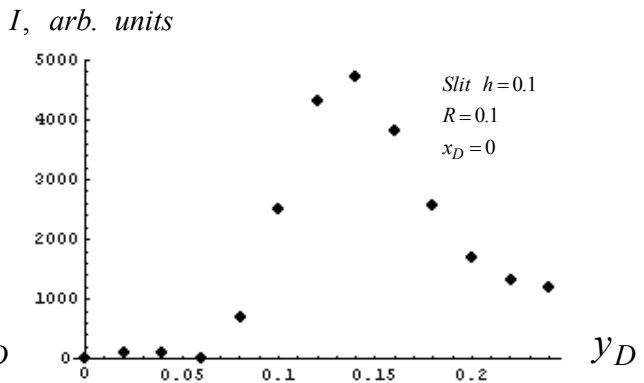

Fig. 12. The same image for pre-wave zone ($R = 0.1$).

The slit halfwidth $h = 0.1$ has been chosen rather small.

In Fig. 12 the analogous distribution, but in the pre-wave zone $(R = 0.1)$ is shown. It can be noted that both distributions coincide with the accuracy less than a few percent in a full analogy with a round hole. The maximum width achieves the quantity $\Delta x_D \approx 0.07$ in both cases. The distributions along the slit edge on the detector (the $x_D$- distribution) calculated for the maximum at $y_D = 0.14$ are given in Fig. 13, 14 for $R = 1$ and 0.1 accordingly. In comparison with the intensity distribution "across" the slit, the distributions "along" the slit are too blurred. .

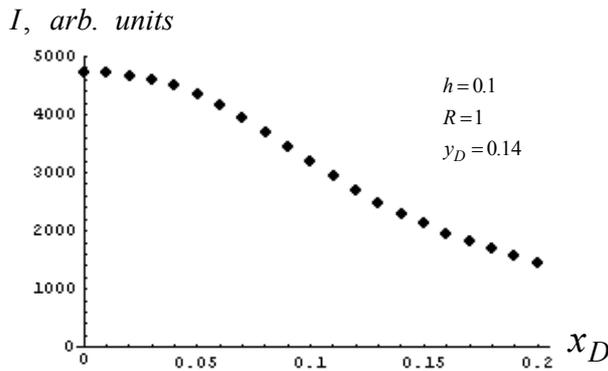
Fig. 13. The intensity distribution of the slit image in wave zone along the slit edge for $y_D = 0.14$.

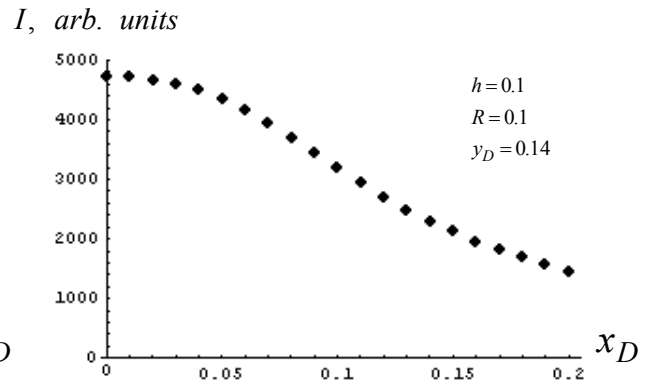
Fig. 14. The same intensity distribution for pre-wave zone ($R = 0.1$).

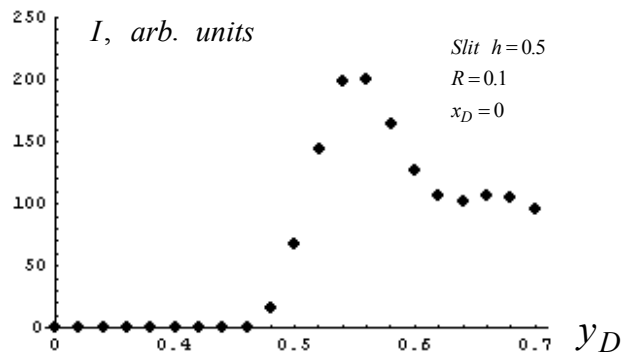
Fig. 15. The image of a "wide" slit ($h = 0.5$) in pre-wave zone ($R = 0.1$).

In Fig. 15 the image of a "wide" slit $(h = 0.5)$ is shown.

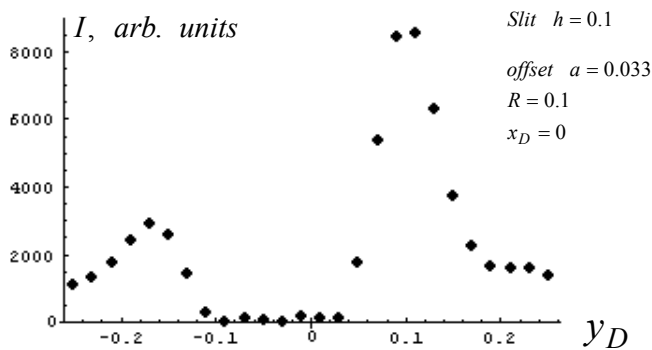
Fig. 16. The image of the "narrow" slit ($h = 0.1$) at asymmetric particle flight ($d = 0.033$, $R = 0.1$).

As in the previous case, a very marked peak near the slit edge is observed, and the characteristic width of the maximum is close to the quantity $\Delta y_D \approx 0.07$.

Fig.11, 13, 15 illustrate the particle flight through the center of the slit. At asymmetric case (in other words, when the distance between the trajectory and the nearest edge of the slit is equal to $h-d$, in that case the distance to the opposite edge is equal to $-h+d$) the slit image in the used coordinate system (the axis $z$ coincides with the particle momentum) will be asymmetric too.

In Fig. 16 the slit image for asymmetric flight ($h = 0.1$; $a = 0.033$) is shown. Different distances to the slit edges (0.067 and 0.133) are fixed on the image clearly.

If the beam flying through the slit has a finite size the shape of the slit image will be distorted. It should be expected that the distortion degree of a perfect distribution "carries" the information about beam size in the same way as the distortion of the angular ODR distribution in wave zone gives the opportunity to get the information about the transverse beam size. It is rather simple to investigate this dependence at approximation $x_m \to \infty$. In this case the function $G_x(x_T, x_D, x_m)$ turns into $\delta$-function:

$$G_{x_L}(x_T, x_D, \infty) = \delta\left(x_T + \frac{x_D}{M}\right), \tag{24}$$

which takes one integration. As before, we shall make calculations for $M = 1$. Thus,

$$\begin{Bmatrix} E_{x_L}^D(x_D, y_D) \\ E_{y_L}^D(x_D, y_2) \end{Bmatrix} = const \int dy_T \begin{Bmatrix} -x_D \\ y_T \end{Bmatrix} \frac{K_1\left(\sqrt{x_D^2 + y_T^2}\right)}{\sqrt{x_D^2 + y_T^2}} \times$$

$$\times exp\left[i\frac{x_D^2 + y_D^2}{4\pi R}\right] G_{y_L}(y_D, y_T, y_m). \tag{25}$$

For the sake of simplicity we shall consider the slit image for a "rectangular" beam, described by the distribution ($y_b$ is a coordinate of a beam particle along $Y_T$-axes).

$$F(y_b) = \begin{cases} \frac{1}{2d}, & -d \leq y_b \leq d \\ 0 & y_b < -d, \ y_b > d \end{cases}$$

because, in this case, the averaging on the beam size may be performed in the simplest way.

Let's consider the beam size effect on the ODR slit image for the SLAC FFTB beam with transverse size $\sim 10$ mcm. In this extremely relativistic case for the wavelength $\lambda = 0.5$ mcm the pre-wave zone parameter $R$ is equal to 0.001 and slit half width $h = 0.003$ looks as desirable (in usual units slit width $2h$ is equal to 28 mcm).

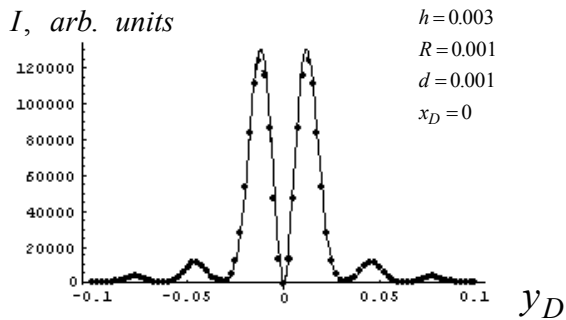
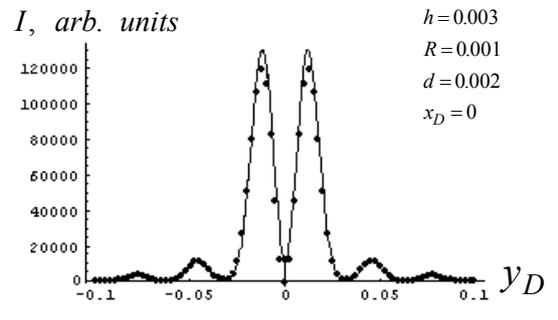

Fig. 17. The narrow slit image in the extremely pre-wave zone ($R = 0.001$; $h = 0.003$) for the finite size beam $-0.001 \leq y_b \leq 0.001$ (points).

Fig.18. The same for the beam with sizes – $0.002 < y_b < 0.002$.

In Fig. 17 the calculated slit image with the beam size $d = 0.001$ (points) in comparison with the perfect image (line) is shown. Results presented on Fig. 18 obtained for two times wider beam ($d = 0.002$).

It should be noticed that the maximum positions don't depend on the beam size, but the radiation intensity ratio between the maximal and minimal values depends on the slit "filling" by the beam. At the increasing of the beam size the particle beam contribution with impact-parameter less than $h$ rises faster for DR from the nearest edge, than the intensity of the radiation from the farthest decreases. As it follows from Fig. 17, 18 the ratio

$$\eta = \frac{I_{min}}{I_{max}},$$

where $I_{min(max)}$ – the intensity of the radiation in the minimum (maximum) is determined by the beam size $d$ and increases with beam size growth (see Fig. 19).

As it follows from the figure, this dependence may be measured experimentally, and it can give the possibility to get the information about beam size.

**6.** The considered geometry for $\gamma = 60000$ and $\lambda = 0.5$ mcm corresponds to the slit width 28 mcm and the beam size $2d = 9$ mcm (Fig.17) and $2d = 18$ mcm (Fig. 18).

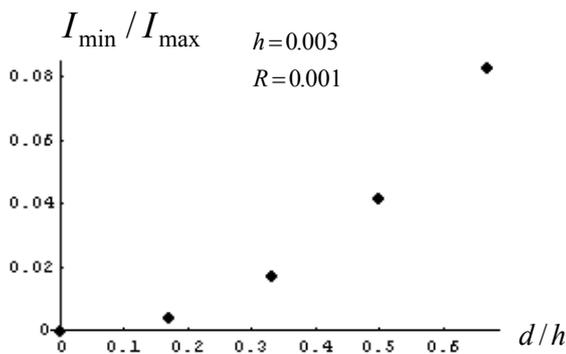

Fig. 19. The dependence of $\eta$ ratio on the beam size.

Extremely pre-wave zone has been considered for $R = 0.001$, that is for the distance between the target and the lens $L = 1.8$ m. To achieve the magnification $\sim 10$ (to obtain the slit image on the detector about 280 mcm, of size that can be easily measured by CCD with pixel size $\sim 20$ mcm) it is necessary to use the short focus lens placed at a short distance.

In conclusion, we can note the following:

i) in the given scheme of ODR beam size monitor the slit width $h$ is determined by the chosen wave length $\lambda$ and the lens aperture $\theta_0$, but doesn't depend on the Lorentz-factor: $h \geq \lambda/\theta_0$;

ii) the higher the method sensitivity is the more the slit filling by the beam is ( the more the ratio $d/h$ is);

iii) the additional focusing of the image along the slit edge increases the image brightness (that is, it gives the possibility to decrease the requirements on sensitivity CCD) at insignificant loss of information..

# References

1. M. Ross, S. Anderson, J. Frisch et al. SLAC-PUB-9280, July 2002.
2. V.A. Lebedev. Nucl. Instrum. and Meth. A 372 (1996) 344-348.
3. M. Castellano, V. Verzilov. Phys. Rev. ST-AB, v. 1 (1998) 062801.
4. K. Honkavaara, X.Artru, R. Chehab et al. Particle Accelerators, v. 63 (1999) 147-170.
5. M. Ross, Review of diagnostics for next generation linear accelerator, SLAC-PUB-8826, May 2001.
6. P. Karataev, PhD Thesis, Tokyo Metropolitan University, 2004.
7. V.A. Verzilov. Phys. Letters A 273 (2000) 135-140.
8. A.P. Potylitsyn. Nucl. Instrum. and Meth. B 145 (1998) 169-179.
9. N. Potylitsyna-Kube, X. Artru. Nucl. Instrum. and Meth. B 201 (2003) 172-183.
10. B.M. Bolotovskii, E.A. Galst'yan, Usp. Fiz. Nauk. 170 (2000) 809.
11. A.P. Potylitsyn. Nucl. Instrum. and Meth. A 455 (2000) 213-216.